\documentstyle[12pt]{article}
\renewcommand{\theequation}{\thesection.\arabic{equation}}
\newcommand{\beq}{\begin{equation}}
\newcommand{\eeq}{\end{equation}}
\newcommand{\bea}{\begin{eqnarray}}
\newcommand{\eea}{\end{eqnarray}}

\begin{document}
\setcounter{page}{0}
\topmargin 0pt
\oddsidemargin 5mm
\renewcommand{\thefootnote}{\fnsymbol{footnote}}
\newpage
\setcounter{page}{0}
\begin{titlepage}

\begin{flushright}
QMW-PH-95-9\\
{\bf hep-th/9503021}\\
March $2nd$, $1995$
\end{flushright}
\vspace{0.5cm}
\begin{center}
{\Large {\bf Conformal Points and Duality of   Non-Abelian Thirring
Models  and
   Interacting WZNW Models}} \\
\vspace{1.8cm}
\vspace{0.5cm}
{\large Chris Hull\footnote{e-mail: C.M.Hull@QMW.AC.UK} and Oleg A.
Soloviev
\footnote{e-mail: O.A.Soloviev@QMW.AC.UK}}\\
\vspace{0.5cm}
{\em Physics Department, Queen Mary and Westfield College, \\
Mile End Road, London E1 4NS, United Kingdom}\\
\vspace{0.5cm}
\renewcommand{\thefootnote}{\arabic{footnote}}
\setcounter{footnote}{0}
\begin{abstract}
{We show that the strong coupling phase of
the non-Abelian Thirring model is dual to the weak-coupling phase of
a system of two  WZNW models  coupled to each
other through a  current-current interaction.
This latter system is integrable and  is related to a perturbed
conformal field theory
which, in the
 large
$|k|$ limit,
has a nontrivial zero of the
 perturbation-parameter beta-function.
The non-Abelian Thirring model reduces to a free fermion theory
plus a topological field theory at this
critical point,
which should therefore be  identified with the isoscalar
Dashen-Frishman conformal point. The
relationship with the Gross-Neveu model is discussed. }
\end{abstract}
\vspace{0.5cm}
\centerline{March 1995}
 \end{center}
\end{titlepage}
\newpage
\section{Introduction}

Gauged WZNW models \cite{KPS},\cite{Hwang}
provide a Lagrangian description of the GKO coset
construction and hence of the minimal conformal models describing
many
statistical systems
\cite{Belavin},\cite{Friedan}, and of many
exact string backgrounds
\cite{Witten1}-\cite{Bardakci1}. These examples
indicate the
importance of Lagrangian CFT's formulated in terms of gauged WZNW
models.
An important observation is that gauged WZNW models  can be
  understood in terms of two ordinary WZNW theories, with Kac-Moody
currents $J_1, \bar J_1$ and
$J_2, \bar J_2$ respectively,
\cite{Novikov}-\cite{Knizhnik} coupled to each other through the
isoscalar Thirring-like
current-current interaction $\lambda J_1^a \bar J_2^a$ at a
particular
value of the Thirring coupling
constant $\lambda$ \cite{Wiegmann}. However, this isoscalar coupling
can be generalised
to renormalizable current-current interactions of the form
$S_{ab} J_1^a \bar J_2^b$ where $S_{ab}$ is a constant matrix
and
this raises the possibility that there may be other
   more
general Thirring current-current interactions that give rise to
new CFT's at special
values of the Thirring coupling constants $S_{ab}$. The fact that
 WZNW models with completely general current-current interactions
emerge
naturally within the coadjoint orbit method \cite{Rai} as well as in
the theory
of chiral WZNW models \cite{Bellucci}-\cite{Gates} give further hints
that this might be the case.
Interacting WZNW models arise also in the  non-abelian
bosonization of ordinary fermionic non-Abelian Thirring models with a
generic current-current
interaction \cite{Wiegmann},\cite{Gonzales}-\cite{Soloviev1}.

Recently, one of us has discussed the conformal symmetry of
non-Abelian
Thirring models at the
so-called Dashen-Frishman conformal points
\cite{Soloviev1},\cite{Soloviev2} which are
generalizations of the isoscalar nonperturbative conformal point
discovered by Dashen and Frishman
\cite{Dashen}.  The conformal symmetry at the Dashen-Frishman
conformal
points can be demonstrated
within the Hamiltonian current-current formalism
\cite{Dashen},\cite{Soloviev2}. At the same time an
attempt to associate these conformal points with zeros of the beta
functions was not successful. It was suggested in \cite{Mitter} that
the
isoscalar Dashen-Frishman conformal point is related to the
strong-coupling
phase of the theory, but no computations were carried out. For
example, one  problem in establishing this was pointed out a
 long time ago by Gross and Neveu (see footnote in
\cite{Gross}). At the classical level the isoscalar Thirring model is
equivalent to the Gross-Neveu
theory after a Fierz rearrangement. A puzzle is that Gross and Neveu
have shown that their model
does not allow nontrivial conformal points to exist, while Dashen and
Frishman have exhibited that
their model does have a nontrivial conformal point.
This suggests that the Gross-Neveu and the Dashen-Frishman models are
different quantizations of
the same classical model.
However, it is possible that the Gross-Neveu and the Dashen-Frishman
models may describe different
phases of the same Thirring model, a weak coupling phase and a strong
coupling phase
respectively. Indeed, Fierz transformations which are allowed in the
Gross-Neveu phase may be
prohibited at the Dashen-Frishman conformal points. At these points
different four fermionic
combinations have different conformal dimensions, so that they cannot
be manipulated using  Fierz
transformations. One of the aims of this paper is to investigate this
possibility further.

Recently, the isoscalar  conformal points of the non-Abelian Thirring
model  were found by the
perturbative method in the large $N$ limit, where $N$ is the number
of
flavours. However, the
method employed in \cite{Bardakci} does not appeal to beta functions
but rather to current algebra.
In the present paper, we will investigate  the conformal invariance
of
the Thirring model considered
in
\cite{Bardakci}  by studying the renormalization group flows.

We will show that the beta function approach leads  to find
nontrivial
critical points of the
fermionic Thirring model.
``Bosonizing" the non-abelian Thirring model gives   free fermions
plus a
system of two
interacting WZNW models at negative level $k$.
We will show that this interacting WZNW model in turn is related to
  a certain perturbed CFT which was introduced in
\cite{Soloviev3},\cite{Soloviev4} (see also
\cite{Soloviev5}) and shown to have a nontrivial infra-red conformal
point.
Thus the system of interacting WZNW models and the Thirring model
also
have such a fixed point,
and
 in the
fermionic Thirring model this conformal point will be argued to
correspond to a free fermion
phase, and so is a generalisation of the Dashen-Frishman isoscalar
conformal point.
Although the WZNW model at negative level  has negative norm states
and
so is non-unitary, we will be interested in a subsector of a unitary
theory. We
will
see that at the critical point the  system of two
interacting WZNW models at negative level becomes the bosonic part of
a
topological field theory  in the large
$|k|$ limit, so that the negative-norm states can be removed by
introducing ghosts and restricting
physical states to be those in the cohomology classes of the
topological field theory BRST operator.

The paper is organized as follows. In section 2, two interacting
level
$k$ WZNW models  are introduced and in section 3 their action   is
re-expressed
 in a factorized form
which will be convenient for implementing the $1/k$ method. In
section
4 we
will consider a perturbation of a  WZNW model by a  relevant operator
that is
intimately related to the models of sections 2,3.
  In section 5 we will discuss the conformal symmetry
of the fermionic Thirring model at the conformal point found for the
bosonic system. We conclude
with some comments on the results obtained.

\section{Interacting WZNW models}

Let $S_{WZNW}(g_1,k_1)$ and $S_{WZNW}(g_2,k_2)$ be the actions of two
 WZNW models of levels  $k_1$,  $k_2$ and with $g_1, g_2$ taking
values
in the
groups $G_1, G_2$ respectively:
\begin{eqnarray}
S_{WZNW}(g_1,k_1)
&=&{-k_1\over4\pi}\left\{\int\mbox{Tr}|g^{-1}_1\mbox{d}g_1|^2~+~\frac{
i}
{3}\int\mbox{d}^{-1}\mbox{Tr}(g_1^{-1}\mbox{d}g_1)^3\right\},\nonumber
\\
\label{wznw} \\
S_{WZNW}(g_2,k_2)
&=&{-k_2\over4\pi}\left\{\int\mbox{Tr}|g^{-1}_2\mbox{d}g_2|^2~+~\frac{
i}
{3}\int\mbox{d}^{-1}\mbox{Tr}(g_2^{-1}\mbox{d}g_2)^3\right\}.\nonumber
\end{eqnarray}
We shall  add the following interaction term
\begin{equation}
S_I={-k_1 k_2\over\pi}\int d^2z~\mbox{Tr}^2(g_1^{-1}\partial
g_1~S~\bar\partial
g_2g_2^{-1})\label{interaction},\end{equation}
to obtain\footnote{A generalization to three and more
interacting WZNW models is straightforward.}
\begin{equation}
S(g_1,g_2,S)=S_{WZNW}(g_1,k_1)~+~S_{WZNW}(g_2,k_2)~+~S_I(g_1,g_2,S)
\label{action},
\end{equation}
with the coupling $S$ belonging to the direct product of two Lie
algebras
${\cal G}_1\otimes{\cal G}_2$.
 The symbol
$\mbox{Tr}^2$ indicates a double tracing over the indices of a matrix
from the
tensor product ${\cal G}_1\otimes{\cal G}_2$.
We shall assume that $S$ is invertible, so that $G_1$ and $G_2$ have
the same dimension, and
 we shall later restrict  ourselves to the case $G_1=G_2=G$.

The
 coupling
matrix $S$ is dimensionless
and the theory described by eq. (\ref{action}) is    conformally
invariant
classically.
If $S=0$ and there is no interaction, the theory has
$\hat G_1^L \times \hat G_1^R\times \hat G_2^L \times \hat G_2^R $
affine symmetry under which
\begin{equation}
g_1\to\bar\Omega_1(\bar z)g_1\Omega_1(z),~~~~~~~~
g_2\to \bar\Omega_2(\bar z)g_2\Omega_2(z).\end{equation}
The parameters $\Omega_{1,2}$ and $\bar\Omega_{1,2}$ are
 arbitrary independent Lie-group-valued
functions of
$z$ and $\bar z$ respectively.
Remarkably, when $S \ne 0$, the interacting theory still has
$\hat G_1^L \times \hat G_1^R\times \hat G_2^L \times \hat G_2^R $
affine symmetry \cite{Soloviev6}, under which
\begin{equation}
g_1\to\bar\Omega_1(\bar z)g_1h_1\Omega_1(z)h_1^{-1},~~~~~~~~
g_2\to h_2^{-1}\bar\Omega_2(\bar
z)h_2g_2\Omega_2(z),\label{symmetry}\end{equation}
where  $h_1,~h_2$ are non-local functions of $g_1,g_2$ satisfying
\begin{equation}
\bar\partial h_1h_1^{-1}=2k_2\mbox{Tr}~S~\bar\partial
g_2g_2^{-1},~~~~~
h_2^{-1}\partial h_2=2k_1\mbox{Tr}~S~g_1^{-1}\partial
g_1.\label{h's}\end{equation}
The $\bar \Omega _1$ and $\Omega_2$ transformations remain local,
while
the $\bar \Omega _2$ and $\Omega_1$ transformations are now
intrinsically non-local, as they
involve $h_1, h_2$.

The local $\bar \Omega _1$ and $\Omega _2 $ symmetries are manifest,
but the  remaining non-local ones
 are by no means obvious, so we now give a direct proof of the
$\Omega_1$ symmetry; an alternative
way of understanding the symmetry is given in Appendix B. In
infinitesimal form
the $\Omega_1$ symmetry is given as follows
\begin{equation}
g_1\to g_1\Lambda_1,\label{Lambda}\end{equation}
where
\begin{equation}
\Lambda_1=h_1\Omega_1h^{-1}_1,~~~~~~~~\Omega_1=1+\epsilon_1,~~~~~~~~~
\bar\partial\epsilon_1=0.\label{epsilon}\end{equation}
The variation of the  WZNW action for $g_1$ is
\begin{equation}
\delta S_{WZNW}(g_1)=-{k_1\over2\pi}\int
d^2z\mbox{Tr}(g^{-1}_1\partial
g_1\bar\partial h_1h_1^{-1}\lambda_1~-~g^{-1}_1\partial
g_1\lambda_1\bar\partial h_1h_1^{-1}),\label{deltawznw}\end{equation}
where
\begin{equation}
\lambda_1=h_1\epsilon_1h^{-1}_1.\label{lambda}\end{equation}
Further, using the definition of $h_1$ given by eq. (\ref{h's}) we
obtain
\begin{equation}
\delta S_{WZNW}(g_1)=-{k_1k_2\over\pi}\int
d^2z\mbox{Tr}^2(\lambda_1g^{-1}_1\partial g_1S\bar\partial
g_2g_2^{-1}~-~g^{-1}_1\partial g_1\lambda_1S\bar\partial
g_2g_2^{-1}).\label{dwznw}\end{equation}
The variation of the interaction term (\ref{interaction}) under
(2.7), (2.8) is
\begin{equation}
\delta S_I=-{k_1k_2\over\pi}\int
d^2z\mbox{Tr}(-\lambda_1g^{-1}_1\partial
g_1S\bar\partial g_2g^{-1}_2~+~g^{-1}_1\partial
g_1\lambda_1S\bar\partial
g_2g^{-1}_2~+~\partial\lambda_1S\bar\partial
g_2g^{-1}_2).\label{dint}\end{equation}
Adding together eq. (\ref{dwznw}) and eq. (\ref{dint}), we finally
obtain
\begin{equation}
\delta S(g_1,g_2,S)=-{k_1k_2\over\pi}\int
d^2z\mbox{Tr}^2\partial\lambda_1S\bar\partial
g_2g^{-1}_2.\label{part}\end{equation}
Using (2.6), the expression on the right hand side of eq.
(\ref{part}) can be
rewritten in terms of $h_1$ as follows
\begin{eqnarray}
\delta S(g_1,g_2,S)&=&-{k_1\over2\pi}\int
d^2z\mbox{Tr}\partial\lambda_1\bar\partial
h_1h^{-1}_1={k_1\over2\pi}\int
d^2z\mbox{Tr}\lambda_1\partial(\bar\partial h_1h^{-1}_1)\nonumber\\
&\label{ds}& \\
&=&{k_1\over2\pi}\int
d^2z\mbox{Tr}\lambda_1h_1\bar\partial(h^{-1}_1\partial
h_1)h^{-1}_1=-{k_1\over2\pi}\int
d^2z\mbox{Tr}\bar\partial(h^{-1}_1\lambda_1h_1)h^{-1}_1\partial
h_1=0,\nonumber\end{eqnarray}
because $h^{-1}_1\lambda_1h_1=\epsilon_1$, which is holomorphic
(2.8). Thus the
action (2.3) is invariant, up to a surface term, under (2.7), (2.8),
and the
$\bar\Omega_2$ symmetry of the action in eq. (\ref{action}) can be
proven in
the similar fashion.

The equations of motion of the theory (\ref{action})
can be written as
\begin{equation}
\partial\bar{\cal J}_1=0,~~~~~~~~\bar\partial{\cal
J}_2=0,\end{equation}
where
\begin{eqnarray}
\bar{\cal J}^{\bar a}_1&=&\bar J^{\bar a}_1~+~2k_2\phi_1^{a\bar
a}~S^{a\bar b}~\bar
J^{\bar b}_2,\nonumber\\ &\label{currents} & \\
{\cal J}^a_2&=& J^ a_2~+~2k_1\phi_2^{a\bar a}~S^{b\bar a}~
J^b_1.\nonumber\end{eqnarray}
and
\begin{eqnarray}
J^a_1&=&-\frac{k}{2}\mbox{Tr}(g_1^{-1}\partial g_1t^a_1),\nonumber\\
J^a_2&=&-\frac{k}{2}\mbox{Tr}(g_2^{-1}\partial g_2t^a_2),\nonumber\\
\bar J_1^{\bar a}&=&-\frac{k}{2}\mbox{Tr}(\bar\partial
g_1g_1^{-1}t^{\bar a}_1),\\
\bar J_2^{\bar a}&=&-\frac{k}{2}\mbox{Tr}(\bar\partial
g_2g_2^{-1}t_2^{\bar
a}),\nonumber\\
\phi_1^{a\bar a}&=&\mbox{Tr}(g_1t_1^ag_1^{-1}t_1^{\bar
a}),\nonumber\\
\phi_2^{a\bar a}&=&\mbox{Tr}(g_2t_2^ag_2^{-1}t_2^{\bar
a}),\nonumber\end{eqnarray}
where $t^a_{1,2}$ are the generators of the Lie algebras ${\cal
G}_{1,2}$ associated with
the Lie groups $G_{1,2}$,
\begin{eqnarray}
\left[t^a_{1},t^b_1\right]&=&f^{ab}_{(1)~c}t^c_1,\nonumber\\ & & \\
\left[t^a_{2},t^b_2\right]&=&f^{ab}_{(2)~c}t^c_2,\nonumber\end{eqnarray}
with $f^{ab}_{(1,2)~c}$ the structure constants.
The local conserved currents $\bar{\cal J}_1, {\cal J}_2$ correspond
to
the
local
 $\bar \Omega _1$ and $\Omega_2$ symmetries,
and there are in addition two non-local conserved currents ${\cal
J}_1,
\bar{\cal J}_2$
corresponding to the non-local $\bar \Omega _2$ and $\Omega_1$
symmetries.
These two local and two non-local conserved currents
generate an infinite number of conserved charges that lead to  the
integrability of the system of
two interacting WZNW models with {\it arbitrary} invertible coupling
matrix $S$ (at least
classically).
The classical current algebras generated by ${\cal J}_1^a$ and
$\bar{\cal J}^{\bar a}_1$ in
eqs. (\ref{currents}) have
 levels $k_2$ and $k_1$ respectively, but we shall see in section 5
that there
is a
conformal point at which these
become renormalized  to
$-k_2-2c_V(G_2)$ and $-k_1-2c_V(G_1)$ respectively.
The current algebras can in principle be used to solve the model; we
hope to return to
this   in the future, and here adopt a different strategy.

{}From now on, we shall take $G_1=G_2=G$, $k_1=k_2=k$.
WZNW models are usually considered for compact groups and positive
integral
level.  In this paper we shall also be interested in WZNW models at
negative (integer) $k$ with compact
$G$. The WZNW model on compact $G$
at negative level is a
complicated system   whose spectrum   contains
 states of negative norm. We will discuss in
section 5 an important situation in which nonunitary states  of WZNW
models at negative level
decouple and can be dropped to leave a unitary theory.

 The classical conformal invariance of the  the system (\ref{action})
will in
general be
 spoiled
by quantum conformal anomalies.  This can be seen in the simple
example
in which
$G=SU(2)$, $k=1$ and $S=\lambda~I$, where $I$ is the identity  in
$su(2)\otimes su(2)$.
When $\lambda$ is small, the theory given by (\ref{action}) is
equivalent to
the
sine-Gordon model \cite{Soloviev7} which is not  a conformal theory
for
general values of the
coupling. However, as for the Sine-Gordon model, there can be special
values of the coupling for which
the conformal invariance extends to the quantum level. For
example, if we  take $ \lambda=1/2k$ so that $S=I/2k$, then using the
Polyakov-Wiegmann formula
\cite{Wiegmann} we obtain
\begin{equation}
S(g_1,g_2,S=I/2k)=S_{WZNW}(f,k),\label{f}\end{equation}
where $f=g_1g_2$. The conformal invariance of this theory follows
immediately
from the fact that the right hand side of eq. (\ref{f}) is a
conformal WZNW
model.
This Polyakov-Wiegmann conformal point
has an extra gauge invariance under $g_1\to
g_1\Lambda,~g_2\to\Lambda^{-1}g_2$, where $\Lambda$ is arbitrary
group
element. Quantization of the theory at the Polyakov-Wiegmann
conformal point
requires gauge fixing in the usual way. The gauge symmetry acts
algebraically
(without derivatives) on $g_1$ and $g_2$, so that with an algebraic
choice such
as $g_2=1$ there are no dynamical ghosts and the theory reduces to a
conventional WZNW model.
Thus (\ref{action}) has at least one conformal point, and one of the
aims of
this paper will be to find
others.

\section{Expansion in the coupling constants}

We shall consider  coupling matrices $S$ of the form
\begin{equation}
S=\sigma\cdot\hat S,\end{equation}
where $\hat S$ is some matrix and $\sigma$ is a small parameter. The
interaction
term in eq. (\ref{action}) is linear in $\sigma$, but we will recast
this
theory
in
a new form without interaction between $g_1$ and $g_2$ but
 with a  nonlinear dependence
on
$\sigma$ instead.

To this end, we make  the following change of variables
\begin{eqnarray}
g_1&\to&\tilde g_1,\nonumber\\ &\label{change} & \\
g_2&\to&h(\tilde g_1)\cdot \tilde g_2,\nonumber\end{eqnarray}
where the function $h(\tilde g_1)$ is the solution of the following
equation
\begin{equation}
\partial h\cdot h^{-1}=-2k\sigma~\mbox{Tr}~\hat S\tilde
g_1^{-1}\partial \tilde
g_1.\label{h}\end{equation}
This determines $h$ up to changes of the form
\begin{equation}
h\to h~\Lambda (\bar z),\end{equation}
where $\Lambda$ is an antiholomorphic matrix function,
$\partial\Lambda=0$,
and we will pick some particular solution $h_0(z,\bar z)$.

Writing
  the action given by eq. (\ref{action}) in terms of the new
variables $\tilde
g_1,~ \tilde g_2$ and using the Polyakov-Wiegmann formula, we obtain
\begin{eqnarray}
S(g_1,g_2,k)&\to& S(\tilde g_1,\tilde g_2,k)
=S_{WZNW}(\tilde g_2,k)~+~S_{WZNW}(\tilde g_1,k)\nonumber\\ & &\\
&+&S_{WZNW}(h_0,k)
{}~-~\frac{k^2\sigma}{\pi}~\int
d^2z~\mbox{Tr}^2~\tilde g_1^{-1}\partial \tilde g_1~\hat
S~\bar\partial
h_0h_0^{-1},\nonumber\end{eqnarray}
where $h_0$ is a   non-local function of $\tilde g_1$ satisfying
(\ref{h}).
Remarkably, after this change of
variables, the   field $\tilde g_2$ completely decouples from
$\tilde g_1$. Note that the Jacobian of
the change of variables in eq. (\ref{change}) is formally equal to
one.

The price we pay for the factorization is a highly nonlocal theory
for
the variable $\tilde g_1$.
While  $\tilde g_2$ is governed simply by a WZNW action, the action
for
$\tilde g_1$
 is
\begin{equation}
S(\tilde g_1)=S_{WZNW}(\tilde g_1,k)~+~S_{WZNW}(h_0(\tilde g_1),k)
{}~-~\frac{k^2\sigma}{\pi}~\int
d^2z~\mbox{Tr}^2~\tilde g_1^{-1}\partial \tilde g_1~\hat
S~\bar\partial
h_0(\tilde
g_1)h_0^{-1}(\tilde g_1),\label{theory}\end{equation}
which is nonlocal as $h_0$ is a non-local function of
$\tilde g_1$.
By considering
$\bar \partial (\partial h\cdot h^{-1})$ and using
eq. (\ref{h}), we obtain
\begin{eqnarray}
&&\bar\partial h_0h_0^{-1}(z,\bar z)=-2k\sigma~\mbox{Tr}~\hat S\tilde
g_1^{-1}(z,\bar
z)\bar\partial_{\bar z}\tilde g_1(z,\bar z)
\nonumber\\
&+&2k\sigma~\int d^2y~\bar\partial_{\bar
z}G(z,\bar z;y, \bar y)~
\upsilon (y, \bar y),\label{h0}
\end{eqnarray}
where
\begin{eqnarray}
\upsilon (y, \bar y)&=&
\mbox{Tr}~\hat S\tilde g_1^{-1}(y,\bar
y)[\bar\partial_{\bar y}\tilde g_1(y,\bar y)
\tilde g_1^{-1}(y,\bar
y),\partial_y\tilde g_1(y,\bar y)\tilde g^{-1}_1(y,\bar y)]\tilde
g_1(y,\bar y)\nonumber\\
&+&[\mbox{Tr}~\hat S\tilde g_1^{-1}(y,\bar
y)\partial_y\tilde g_1(y,\bar y),\bar\partial h_0(y,\bar
y)h_0^{-1}(y,\bar
y)]\end{eqnarray}
and the Green function $G(z,\bar z;y, \bar y)$ satisfies
\begin{equation}
\bar\partial_{\bar z}\partial_z G(z,\bar z;y, \bar y)=\delta
(z,y)\delta (\bar
z,\bar y).\end{equation}
We regularise the Green function in such a way   that
\begin{eqnarray}
\lim_{y\to z, \bar y \to \bar z}\partial_zG(z,\bar z;y, \bar
y)=0\nonumber\end{eqnarray}
so that, despite its nonlocality, the right
hand side of eq. (\ref{h0}) is well defined, even when   $(y,\bar
y)\to
(z,\bar z)$. Note that eq. (3.26) fixes
the dependence of $h_0$ on $\bar z$.

In the Wess-Zumino term in $S_{WZNW}(h_0(\tilde g_1),k)$, $h_0$ and
$\tilde g_1$ are extended to
functions of
$z,\bar z$ and an extra coordinate $t$ and an equation analogous to
(\ref{h0}) can be found for $\partial _t h_0 h_0^{-1}$. Then $h_0$
appears
in (\ref{theory}) only through its
derivatives, so that
(\ref{theory}) can be written in terms of $\tilde g_1$ using
(\ref{h}),
(\ref{h0}) and
the analogous equation for
$\partial _t h_0 h_0^{-1}$.
The resulting action includes a non-local term of order $\sigma ^2$
of
the form
\begin{equation}
{k\sigma ^2 \over\pi}  ~\int d^2z ~\bar\partial(\tilde
g^{-1}_1\partial\tilde g_1) \Psi \end{equation}
where
\begin{equation}
\Psi(z,\bar z) = k^2 ~\int d^2y~ G(z,\bar z;y, \bar y)\mbox{Tr}\hat S
\upsilon(y,\bar y) .
\end{equation}
On making a further change of variables
\begin{equation}
\tilde g_1 \to \tilde g_1 (1 + 2\sigma ^2 \Psi + {\cal O}(\sigma ^3))
\end{equation}
we obtain an action which is local up to order $\sigma ^3$ and which
takes the simple form
\begin{equation} S(\tilde g_1)=S_{WZNW}(\tilde
g_1,k)~+~\frac{k^3\sigma^2}{\pi}~\int
d^2z~\mbox{Tr}\left(\mbox{Tr}~\hat S\tilde g_1^{-1}\partial \tilde
g_1\cdot\mbox{Tr}~\hat S\tilde
g_1^{-1}\bar\partial \tilde g_1\right)~+~{\cal
O}(\sigma^3).\label{eq}\end{equation}
Further,  eq. (\ref{eq}) can be rewritten
as
\begin{equation} S(\tilde g_1)=S_{WZNW}(\tilde g_1,k)~-~\tau~\int
d^2z~\Sigma_{ab}\tilde
J_1^a\bar{\tilde J_1^{\bar b}}\tilde\phi_1^{b\bar b}~+~{\cal
O}(\sigma^3),\label{eq1}\end{equation}
where we have introduced the   notation
\begin{equation}
\tau=-\frac{4k\sigma^2}{\pi},~~~~~~~~~
\Sigma_{ab}=\hat S^{a\bar a}\hat S^{b\bar
a}.\end{equation}

The expression for $S(\tilde g_1)$ obtained in eq. (\ref{eq1})
presents an
expansion in the
coupling $\sigma$ around the conformal theory described by the
 level $k$ WZNW model $S_{WZNW}(\tilde
g_1,k)$. In the next section we will show that the operator
$\Sigma_{ab}\tilde J_1^a\tilde{\bar
J_1^{\bar b}}\tilde\phi_1^{b\bar b}$ provides a relevant
renormalizable
perturbation of the WZNW
model at negative level $k$.

\section{Relevant perturbations}

In this section we are going to discuss a certain perturbed CFT which
will be shown to coincide with
the theory given by eq. (\ref{eq1}), to lowest order in the
perturbation
parameter.

Consider
a  WZNW model on group $G$ with level $l$. We
define the following composite operator \cite{Soloviev3} -
\cite{Soloviev5}
\begin{equation}
O^{L,\bar L}=L_{ab}\bar L_{\bar a\bar b}~:J^a\bar J^{\bar
a}\phi^{b\bar
b}:,\label{operator}\end{equation}
where
\begin{eqnarray}
J&=&J^at^a=-\frac{l}{2}g^{-1}\partial g,\nonumber\\
\bar J&=&\bar J^at^a=-\frac{l}{2}\bar\partial gg^{-1},\\
\phi^{a\bar a}&=&\mbox{Tr}:g^{-1}t^agt^{\bar
a}:.\nonumber\end{eqnarray}
The product of the three operators in eq. (\ref{operator}) is defined
by
\cite{Soloviev3}
\begin{equation}
O^{L,\bar L}(z,\bar z)=L_{ab}\bar L_{\bar a\bar
b}~\oint\frac{dw}{2\pi
i}\oint
\frac{d\bar w}{2\pi i}{J^a(w)\bar J^{\bar a}(\bar w)\phi^{b\bar
b}(z,\bar
z)\over|z-w|^2},\label{normal}\end{equation}
where the product in the numerator of the integrand is to be
understood
as an OPE. It
is easy to see that the given product does not contain singular terms
provided
the matrices $L_{ab}$ and $\bar L_{\bar a\bar b}$ are symmetric
\footnote{Indeed, the field $\phi$ is an affine primary vector.
Therefore, its
OPE with the affine current $J$ is
\begin{eqnarray}
J^a(w)\phi^{b\bar b}(z,\bar z)={f^{abc}\over(w-z)}\phi^{c\bar
b}(z,\bar
z)~+~regular~terms.\nonumber\end{eqnarray}
Substituting this formula into eq. (\ref{normal}), one can see that
only
regular terms
will contribute provided $L_{ab}$ is a symmetric matrix.}.

The operator $O^{L,\bar L}$ is a level one affine descendant of the
affine primary field $\phi$.
Indeed, $O^{L,\bar L}$ can be presented in the form
\begin{equation}
O^{L,\bar L}(0)=L_{ab}\bar L_{\bar a\bar b}~J^a_{-1}\bar J^{\bar
a}_{-1}\phi^{b\bar b}(0),\end{equation}
where
\begin{equation}
J_m^a=\oint\frac{dw}{2\pi i}~w^mJ^a(w),~~~~~~\bar
J_m^a=\oint\frac{d\bar w}
{2\pi i}~\bar w^m\bar J^a(\bar w).\end{equation}
Being an affine descendant, the operator $O^{L,\bar L}$ is a
Virasoro primary operator. Indeed, one can check that the state
$O^{L,\bar
L}(0)|0\rangle$ is a highest weight vector of the Virasoro algebra,
with
$|0\rangle$ the $SL(2,C)$-invariant vacuum. That is,
\begin{equation}
L_0~O^{L,\bar L}(0)|0\rangle=\Delta_O~O^{L,\bar L}(0)|0\rangle,~~~~~~
L_{m>0}~O^{L,\bar L}(0)|0\rangle =0.\end{equation}
Here $L_n$ are the generators
of the Virasoro algebra associated with the affine-Sugawara
stress-energy tensor of the conformal
WZNW model.  By using the Knizhnik-Zamolodchikov formula for
anomalous
conformal dimensions
\cite{Knizhnik}, it is straightforward to show that
\begin{equation}
\Delta_O=\bar\Delta_O=1~+~{c_V\over l+c_V}.\label{dim}\end{equation}
Here $\bar\Delta_O$ is the conformal dimension of $O^{L,\bar L}$
associated
with antiholomorphic conformal transformations. The quantity $c_V$ is
defined
by
\begin{equation}
f^{ac}_df^{bd}_c=-c_V~g^{ab}\end{equation}
with $g^{ab}$ the Killing metric.

{}From the formula (\ref{dim}) for anomalous conformal dimensions of
the
operator $O^{L,\bar
L}$, it is clear that when
\begin{equation}
l<-2c_V,\label{cond}
\end{equation}
 the conformal dimensions are
in the range between 0 and 1. Hence, when condition (\ref{cond}) is
fulfilled,
the operator $O^{L,\bar L}$ becomes a relevant conformal operator.
Correspondingly, for positive $l$ the operator $O^{L,\bar L}$ is
irrelevant.

We are interested here in relevant operators, so we shall consider
levels  $l$ which are negative
integers. The conformal WZNW model with negative level is a
nonunitary
theory because there are
 states of negative norm in its spectrum. Let us show that the
operator
$O^{L,\bar L}$ corresponds
to a unitary highest weight representation of the Virasoro algebra of
the nonunitary WZNW model.
Indeed, as we have pointed out above, $O^{L,\bar L}$ has positive
conformal dimensions. Besides, the
condition (\ref{cond}) guarantees that the Virasoro central charge of
the
nonunitary WZNW model is greater
than one,
\begin{equation}
c_{WZNW}(l)={l\dim G\over l+c_V}=\dim G~+~{\cal
O}(1/l)>1.\end{equation}
Thus, the operator $O^{L,\bar L}$ lies in the unitary range of the
Kac-Kazhdan
determinant and so it provides a unitary representation of the
Virasoro
algebra. Note that although  $O^{L,\bar L}$ belongs to a
nonunitary representation of the affine algebra, the Virasoro
representation generated by $O^{L,\bar L}$ is unitary. This can be
understood
from the point of view of analytic continuation of nonunitary WZNW
models. Indeed, one way of thinking about certain nonunitary WZNW
models is to
treat
them as analytic continuations of unitary WZNW models based on
compact groups
to ones on noncompact groups.  In the course of analytic
continuation, the affine operators become nonhermitian, while the
Virasoro
operators continue to be hermitian. We do not insist on
representations of the
affine algebra since we are interested in scaling properties of the
model.
These properties
 allow us to consider the operator $O^{L,\bar L}$ as a physical
conformal
operator\footnote{In the next section we will show that interacting
nonunitary
WZNW models emerge as the bosonic subsector of a certain unitary
theory. Adding
a ghost sector gives a theory with a unitary physical subsector
defined by the
cohomology of a certain BRST operator.}.

Let us turn to the fusion algebra of $O^{L,\bar L}$.
Clearly, operators $O^{L,\bar L}$ with arbitrary symmetric matrices
$L_{ab},~\bar L_{\bar a\bar b}$ are Virasoro primary vectors with the
same
conformal dimensions. However, their fusion algebras may be
different.
Among
all such operators, there are operators which obey the following
fusion
algebra \cite{Soloviev5}
\begin{equation}
O^{L,\bar L}\cdot
O^{L,\bar L}=\left[O^{L,\bar
L}\right]~+~\left[I\right]~+~...,\label{OPE}\end{equation}
where the square brackets denote the contributions of $O^{L,\bar L}$
and the
identity operator $I$ and their descendants, whereas the dots stand
for
all other
 operators with irrelevant conformal dimensions\footnote{We emphasize
that we
are computing the OPE of  conformal operators of the conformal WZNW
model.} .
These operators
must be
singlets under $G$ which are constructed from $g$ and its
derivatives. From the
results of \cite{Knizhnik} on the spectrum of the WZNW model, it
follows that
for general $L$, the only relevant singlet operators that can appear
on the
right hand side of (4.45) are of the form
$O^{\tilde L,\tilde{\bar L}}$ and $I$, for some symmetric matrices
$\tilde
L,~\tilde{\bar L}$; this is shown in detail for $G=SU(2)$ in appendix
C.
As shown in \cite{Soloviev5}, requiring that $L=\tilde L$ so that the
fusion
algebra is given by eq. (4.45) results in complicated algebraic
equations for
the
matrices  $L,~\bar L$.  In
general, these equations may have many solutions each of which will
yield an operator $O^{L,\bar L}$
satisfying the closed fusion algebra in eq. (\ref{OPE}). In what
follows, we
will be dealing with the
large $|l|$ limit. In this limit there are asymptotic solutions for
$L_{ab},~\bar L_{\bar a\bar b}$
\cite{Soloviev5}
\begin{eqnarray}
L_{ab}&=&{1\over\sqrt c_V}~g_{ab}~+~{\cal O}(1/l),\nonumber\\ & & \\
\bar L_{\bar a\bar b}&=&{1\over\sqrt c_V}~g_{\bar a\bar b}~+~{\cal
O}(1/l).\nonumber
\end{eqnarray}
Now it becomes clear that with the given values of the matrices
$L,~\bar L$ one can
make use of the operator $O^{L,\bar L}$ to perform relevant
renormalizable perturbations of the
level $l$ WZNW model.

The perturbed theory is defined as follows
\begin{equation}
S(\epsilon)=S_{WZNW}(g,l)~-~\epsilon~\int d^2z~O^{L,\bar L}(z,\bar z)
+{\cal O}( \epsilon ^2).\label{perturb}
\end{equation}
This theory has been proven \cite{Soloviev5} to possess a nontrivial
infra-red conformal point at
\begin{equation}
\epsilon=\epsilon^{*}\equiv -{2c_V\over\pi l} ~+~{\cal O}(l^{-2})
\end{equation}
for large $|l|$. In the evaluation of $\epsilon^{*}$, the expansion
of the beta
function in $\epsilon$ up to $\epsilon^3$ terms has been used
\cite{Soloviev3},\cite{Soloviev4}, and this is sufficient for large
$|l|$.
Substituting this value of $\epsilon$ gives the new CFT
\begin{equation}
S(\epsilon^{*})=S_{WZNW}(g,l)~-~\epsilon^{*}~\int d^2z~O^{L,\bar
L}(z,\bar z) +{\cal O}(l^{-2}).\label{new}
\end{equation}
It was shown in \cite{Soloviev3} that this new CFT has, in the large
$|l|$
limit, the same
Virasoro central charge and that the operators have the same
anomalous
conformal dimensions
as for the WZNW model with level $-l$. It was argued in
\cite{Soloviev3}
that the  WZNW model at level $l$ flows to a model which, at the
infra-red
fixed point, is     the WZNW model with  level given approximately by
$-l$ for
large $|l|$. We will argue later that the exact result for  the level
of the
new WZNW theory is $-l-2c_V$.

It is interesting to compare the model just obtained  with the theory
discussed in the
previous section. The model in eq. (\ref{new}) coincides with the
expression
in
eq. (\ref{eq1}) provided $k=l$ and
\begin{equation}
\sigma^2={c_V\over2k^2},~~~~~~~~~\hat S^{a\bar a}\hat S^{b\bar
a}={g^{ab}\over c_V}.
\end{equation}
Thus when these conditions are satisfied,
 the
system of two interacting WZNW models at level $k$ is conformally
invariant. When $|k|$ is large,
there are two solutions
\begin{equation}
S^{*}_{\pm}=\pm{I\over\sqrt2k}.\label{point}
\end{equation}
In fact there are further solutions in the large $|k|$ limit, which
correspond
to other solutions $L_{ab}$ of the algebra in eq. (\ref{OPE})
\cite{Soloviev5}.
However only the two solutions displayed above are nondegenerate, so
that $(S^{*})^{-1}$ exists \cite{Soloviev5}.
The invertibility will be an important property in what follows.
The status of the $S^{*}_+$ solution is
unclear: in flowing from zero coupling to
$S^{*}_+$, one passes through the Polyakov-Wiegmann conformal point
$S^{*}_{PW}=I/2k$,
at which there is some evidence that a phase transition occurs.
This means that one might expect some non-analytic behaviour in
continuing the
coupling constant through the Polyakov-Wiegmann conformal point to
$S^{*}_+$.
 However, it might be the case that the
$S^{*}_+$ solution is related to the $S^{*}_-$ solution by some kind
of duality
transformation.
In the next section, we shall investigate
  the second
solution $S^{*}_-$
using a fermionic version of the theory.

\section{Free fermion phase of the non-Abelian Thirring model}

Interacting WZNW models at
negative level naturally emerge in the course of the path integral
bosonization of an ordinary
non-Abelian fermionic Thirring model
\cite{Wiegmann} and this is the main motivation for the study of the
model (\ref{action}).
Consider the fermionic non-Abelian Thirring model described by the
following action
\begin{equation}
S_F={1\over4\pi}\int d^2z\left(\bar
\psi_L\bar\partial\psi_L~+~\bar\psi_R\partial\psi_R~-~G_{a\bar
a}J^a_LJ^{\bar a}_R\right),\label{thirring}
\end{equation}
where $\psi_L$ and $\psi_R$ are complex Weyl spinors transforming as
the fundamental representations of groups $G_L$ and $G_R$
respectively. The spinors $\psi^i_R$ and $\psi^{\bar i}_L$ carry
flavour indices $i=1,...,k_R$ and $\bar i=1,...,k_L$. The last term
in
eq. (\ref{thirring}) describes the
general interaction between fermionic currents
\begin{equation}
J_L^a=\bar\psi_Lt^a\psi_L,~~~~~ J^{\bar a}_R=\bar\psi_Rt^{\bar
a}\psi_R.
\end{equation}
Here $t^a,~t^{\bar a}$ are the generators in the Lie algebras ${\cal
G}_L,~{\cal G}_R$:
\begin{equation}
[t^a,t^b]=f^{ab}_ct^c,~~~~~~~~
[t^{\bar a},t^{\bar b}]=f^{\bar a\bar b}_{\bar c}t^{\bar c}.
\end{equation}
$G_{a\bar a}$ is a coupling constant matrix. In what follows we will
suppose $G_L=G_R=G,~k_L=k_R=N$.

The  action  (\ref{thirring}) can be rewritten in the equivalent form
\begin{equation}
\tilde S_F={1\over4\pi}\int
d^2z\left(\bar\psi_L\bar\partial\psi_L~+~\bar\psi_R\partial\psi_R~+~A_
aJ
^a_L~+~\bar A_{\bar
a}J^{\bar a}_R~+~(G^{-1})^{a\bar a}A_a\bar A_{\bar
a}\right),\label{gaussian}
\end{equation}
where $A_a,~\bar A_{\bar a}$ are auxiliary vector fields. This
  yields the fermionic Thirring model after eliminating the fields
$A,~\bar A$  by using their
algebraic equations of motion. After introducing these auxiliary
fields, the action
(\ref{gaussian}) takes a Gaussian form for the fermions.

Note that the fields $A,~\bar A$ are not to be treated as gauge
vector
fields. The theory is only gauge invariant if $G^{-1}\sim\hbar I$. In
particular, for general $G$ there is no  gauge
symmetry that
can  be used to fix   local counterterms.
The partition functions  $Z_F(G)$ and $ \tilde Z_F$ derived from the
two forms of the action, (5.52) and
(5.55),  are related by
\begin{equation}
Z_F(G)=J^{-1}~\tilde Z_F,
\end{equation}
where
\begin{eqnarray}
Z_F(G)&=&\int{\cal D}\psi_L{\cal D}\psi_R~\mbox{e}^{-S_F},\nonumber\\
\tilde Z_F&=&\int{\cal D}\psi_L{\cal D}\psi_R{\cal D}A{\cal D}\bar
A~\mbox{e}^{-\tilde S_F},\\
J&=&\int{\cal D}A{\cal D}\bar A~\mbox{e}^{-{1\over4\pi}\int
d^2z~AG^{-1}\bar A}.\nonumber
\end{eqnarray}
The quantity $J$ is an inessential constant which we will drop
(see Appendix A). The functional integrals over the spinor fields are
Gaussian and the
variables $\psi_L$ and $\psi_R$ do not mix. Since there is no   gauge
symmetry to fix the
counterterms, we choose to quantize in such a way as to preserve
holomorphic factorization.
 Then the
computation of the fermion integrals amounts to calculating the
functional determinants of the
chiral differential operators. These determinants can be evaluated by
the Leutwyler method
\cite{Leutwyler}, and in this method the only interaction between $A$
and $\bar A$ is given by
the last term in eq. (\ref{gaussian}). We make the following change
of
variables
\begin{eqnarray}
\bar A&\to&\bar\partial g_Rg^{-1}_R,\nonumber\\ &\label{var} & \\
 A&\to&g^{-1}_L\partial g_L.\nonumber
\end{eqnarray}
and the corresponding Jacobian     can be again computed according to
the Leutwyler method.
This Jacobian is the product of the ``ghost" partition function
\begin{equation}
Z_{ghost}=\int{\cal D}b{\cal D}\bar b{\cal D}c{\cal D}\bar c~\exp
\left[-\int d^2z(b\bar\partial
 c+\bar b\partial\bar c)\right],\label{ghost}
\end{equation}
where $b,~c,~\bar b,~\bar c$ are
Grassmann odd auxiliary fields in the adjoint representation of
${\cal
G}$, and the partition
function
$Z_B(-N-2c_V,S)$ of a system of two interacting  WZNW models
with action (\ref{action}) and  levels $k_1=k_2=(-N-2c_V)$.
The coupling matrix
$S$ is related to the coupling matrix $G$. However, the explicit form
of this relation depends on
the regularization scheme and corresponds to the possibility of
adding
a local $A\bar A$
counterterm.  With a chiral regularization, the relation between $S$
and $G$ is given by
\begin{equation}
S=-{G^{-1}\over(N+2c_V)^2}.\label{dual}
\end{equation}

We finally  arrive  at the following relation
\begin{equation}
Z_F(G)=Z_F(0)Z_B(-N-2c_V,S)Z_{ghost},\label{partition}
\end{equation}
This implies that conformal points of the system of two interacting
WZNW models are conformal points of the fermionic Thirring model.
The fact that the system of two interacting
WZNW models had
$\hat G_1^L \times \hat G_1^R\times \hat G_2^L \times \hat G_2^R $
affine symmetry implies that the Thirring model should also, and so
should be integrable.
In the right hand side of eq.
(\ref{partition}) we have  the partition function of interacting WZNW
models at
negative level $(-N-2c_V)$.
 For the bosonic system of two interacting WZNW models we
 found a nontrivial conformal point $S^{*}_-$ given by eq.
(\ref{point}) for
large $|k|$.
Since $k=(-N-2c_V)$, we will consider the large $N$ limit.
 Note that the value of $S^{*}_-$ does not
depend on the regularization procedure, so that the theory at the
critical point is determined
unambiguously. From formula (\ref{dual}),  the point $S=0$
corresponds to
$G=\infty$. Hence, the expansion in $S$ or $\sigma$ around free WZNW
models amounts to a strong
coupling expansion of the fermionic Thirring model around
$G=\infty$. Of course, this would cause many problems in the
fermionic
theory, while in the bosonic
system this is a weak coupling regime that can be studied
perturbatively.

In the large $N$ limit, $S_B(-N-2c_V,S)$ is the model (2.3) with
$k\sim -N$,
which can then be re-expressed as (3.24). The $g_1$ dependence is, as
we have
seen, described by a model which has an infra-red fixed point at
$S_-^{*}$ at
which the theory is a WZNW model of level given (for large $N$)
approximately
by $+N$. We then find
\begin{equation}
S(g_1,g_2,S^{*})=S_{WZNW}(g_1,N)~+~S_{WZNW}(g_2,-N).
\end{equation}
 As a result, the relation
(\ref{partition})
 takes the form
\begin{equation}
Z_F(G^{*})=Z_F(0)Z_{WZNW}(N)Z_{WZNW}(-N)Z_{ghost}\label{Z}
\end{equation}
in the large $N$ limit.

The partition function of the $G/G$ coset is given by
\begin{equation}
Z_{G/G}=Z_{WZNW}(k)Z_{WZNW}(-k-2c_V)Z_{ghost},
\end{equation}
where $k$ is the level of the WZNW model on $G$ \cite{Spiegelglas}.
With
$k=N$, $Z_{WZNW}(-N-2c_V) \sim Z_{WZNW}(-N)$ in the large $N$ limit,
so
that
 (\ref{Z}) takes
the  following form
\begin{equation}
Z_F(G^{*})=Z_F(0)Z_{G/G}
\end{equation}
in the large $N$ limit.
 Although  $Z_{G/G}$ represents states of both negative norm and
positive norm, it corresponds to a
topological field  theory \cite{Spiegelglas}
and so can be quantized in such a way that, at this conformal point,
the spectrum consists of only ground states, so that  the only
dynamical
degrees of freedom of the full theory at this critical point are
described by
$Z_F(0)$, which is the partition function of the system of free
fermions. At  this conformal point
the (scheme-dependent) value of the Thirring coupling constant is
given
by
\begin{equation}
G^{*}=-{(S^{*})^{-1}\over N^2}=-{\sqrt2\over N}
\end{equation}
to lowest order in $1/N$.
Although the critical value of the Thirring coupling depends on the
regularization
procedure, the effective CFT is known exactly to leading order in
$1/N$, so that the
fermionic non-Abelian Thirring model is effectively a theory of free
fermions at this new conformal
point.
The   conformal point   found by Dashen and Frishman \cite{Dashen}
also
gave
a free fermion theory, suggesting that the conformal point we have
found is precisely the
Dashen-Frishman one.    Dashen and Frishman made use of the
nonperturbative current-current method for analysis of the conformal
symmetry of the Thirring model,
enabling them to find the critical value of $G$ exactly, with the
result
\begin{equation}
G^{*}={1\over N+c_V(G)}.
\end{equation}
 Using this result, we  find the finite shift of the level along the
flow from the non unitary to
unitary WZNW models:
$S_{WZNW}(g,-N-2c_V(G))\to S_{WZNW}(g,N)$. Another interesting
observation   is that the
classical level $-N-2c_V(G)$ of the affine currents ${\cal J}_2^a$
and
$\bar{\cal J}^{\bar a}_1$ in
eqs. (\ref{currents}) gets renormalized at the Dashen-Frishman
conformal point,
becoming equal to $N$.

Note that one can always add the counterterm $A\bar A$ with an
appropriate coefficient to fix the
constant $G^{*}$ at the Dashen-Frishman value. The advantage of the
path integral formalism compared
to the Hamiltonian current-current quantization is that the former
allows to derive the effective
CFT exactly. In the Dashen-Frishman approach free fermions emerge
through the
operator bosonization procedure.

Note that the same effect has been known for a long time  in the
sine-Gordon model. Coleman has proven
that the sine-Gordon model is effectively a theory of free Dirac
spinors at the special value of the
coupling $\beta$ \cite{Coleman}. The sine-Gordon model in its turn
can
be described as the fermionic
Thirring model of $SU(2)$ spinor fields with one flavour
\cite{Soloviev7}. Therefore, the free
fermion regime of the sine-Gordon model results in the free fermion
regime of the fermionic
non-Abelian Thirring model at   particular values of the Thirring
coupling constants. We have
shown that the free fermion phase is a characteristic feature of all
fermionic non-Abelian
Thirring models.

Even away from the conformal points,
the partition function for the fermionic non-Abelian Thirring model,
which is
clearly unitary,
takes the factorised form
\begin{equation}
Z_F(G)=Z_F(0)Z_{WZNW}(-N-2c_V(G))Z_{ghost}Z(S),\label{beyond}
\end{equation}
where $Z(S)$ is the partition function of the theory given by eq.
(\ref{theory}). The first three factors in eq. (\ref{beyond})
correspond to the
partition function of free spinors coupled to gauge fields, so that
their
product gives a unitary partition function. Thus the factor $Z(S)$
must be the
partition function of a unitary theory  by itself .

\section{Conclusion}

In this paper  we have found a  nontrivial
quantum
field theory described by the action in eq. (\ref{theory}) which
possesses
nontrivial conformal points. We
have shown that conformal points of this bosonic theory are related
to
critical points of the fermionic
non-Abelian Thirring model. We have studied in detail one conformal
point of the Thirring model at
which the theory turns out to correspond to an effective theory of
free fermions.
There may be other
 conformal points in these models, corresponding to the
 many relevant perturbation operators $O^L$ constructed from
 different $L$'s.
Indeed, the models considered here might
  be useful in the search for   a   Lagrangian formulation of the
affine-Virasoro construction \cite{Halpern}.

Finally, we return  to the apparent paradox in the relation between
the
Gross-Neveu
model and the
 Dashen-Frishman model. Gross and Neveu quantized    the Thirring
model in the weak coupling regime, perturbing about the
   free fermion theory in terms of a   small coupling $G$.
We have shown that the Dashen-Frishman conformal point is to be found
in a strong coupling
regime of the Thirring model, expanding about
$G=\infty$ (for a qualitative discussion of the given point see
\cite{Mitter}). Although there is no well defined expansion in the
strong
coupling for the fermionic
theory, we found a dual bosonic  formulation in which the
Dashen-Frishman point occurs at weak
coupling. Although  in the
limit
$N\to\infty$ the value of the Dashen-Frishman conformal point is
close
to $G=0$, this point can be
reached only from $G=\infty$ and not from $G=0$. In fact, the $1/N$
expansion around the free fermion
model breaks down when one approches the Dashen-Frishman conformal
point from the weak coupling
regime. There are correlation functions which become singular at the
critical value of the coupling
constant
\cite{Kutasov}, despite the fact that the Gross-Neveu beta function
is
regular at this point.There
is  a strong similarity between the Dashen-Frishman conformal point
and
the Polyakov-Wiegmann
conformal point. This indicates that at the Dashen-Frishman conformal
point the fermionic Thirring
model undergoes a phase transition. Since Gross and Neveu have used
the free fermion theory as the vacuum,
their computations are only valid in the weak-coupling phase around
the
free fermion theory.
Therefore, there is no way to see the Dashen-Frishman critical point
within the  Gross-Neveu
perturbation theory; to find it, non-perturbative or strong-coupling
methods are required.

\par \noindent
{\em Acknowledgements}:

We would like to thank Michael Green and Jonathan Evans for helpful
discussions.
O.S. would  like to thank the PPARC
for financial support.

\bigskip

{\large\bf Appendix A}\vspace{.15in}
\renewcommand{\theequation}{A.\arabic{equation}}
\setcounter{equation}{0}

In this appendix, we give a check on the concistency of our
computations.  Consider the
following functional integral
\begin{equation}
F=\int{\cal D}B{\cal D}\bar B~\exp\left({1\over\pi}\int d^2z~BM\bar
B\right),\end{equation}
where $B,~\bar B$ are vector fields in the adjoint representation of
a
group $G$; $M$ is a
nondegenerate constant matrix. The functional integral $F$ does not
contribute to the
conformal anomaly at any value of $M$, and this will play a crucial
role in our check.

We make the following change of variables:
\begin{equation}
B=g_1^{-1}\partial g_1,~~~~~~~\bar B=\bar\partial
g_2g^{-1}_2,\end{equation}
where $g_1,~g_2$ are $G$-valued fields.
The Jacobians are found in the same way as those for (\ref{var}),
with the
result that $F$ is now given
by
\begin{equation}
F=Z_B\left(-2c_V(G),{M\over4c^2_V(G)}\right)~Z_{ghost},\end{equation}
where $Z_B$ is the partition function of the
 system of two interacting level $(-2c_V)$ WZNW models with the
coupling matrix $M/4c^2_V$;
$Z_{ghost}$ is the partition function of the ghost-like fields (see
eq.
(\ref{ghost})).

Thus, the conformal properties of $F$ are determined
 by the conformal properties of $Z_B$ at  level $(-2c_V)$. The claim
that $F$ does not
contribute to the conformal anomaly amounts to the statement that the
sum of the  Virasoro central
charges for the conformal models on the right hand side of eq. (A.3)
vanishes for all $M$.This is
difficult to verify explicitly in general, but can be checked at the
conformal points we have
found in this paper.

Let us first check the Polyakov-Wiegmann conformal point which
corresponds to
\begin{equation}
M_{PW}=-2c_VI.\end{equation}
At this point the theory acquires  additional gauge symmetry (see
section 1). This symmetry is
sufficient to gauge away one of the two group elements. Therefore, at
the PW critical point, the
quantity $F$ is given by
\begin{equation}
F(M_{PW})=Z_{WZNW}(-2c_V)~Z_{ghost}.\end{equation}
It is easy to see that the Virasoro central charge in the right hand
side of eq. (A.5)
 indeed vanishes:
\begin{equation}
c=c_{WZNW}(-2c_V)~+~c_{ghost}={(-2c_V)\dim G\over-2c_V+c_V}~-~2\dim
G=0.\end{equation}

Now let us turn to the isoscalar Dashen-Frishman conformal point
which
corresponds to
\begin{equation}
M_{DF}=2\sqrt{2}  c_VI.\end{equation}
According to our computations, at the DF fixed point the quantity $F$
is given by
\begin{equation}
F(M_{DF})=Z_{G/G}.\end{equation}
Obviously, the Virasoro central charge in the right hand side of eq.
(A.8) vanishes.

Thus, we have proven that $F(M)$ does not contribute into the
conformal
anomaly at
least for two different values of $M$, checking the correctness of
our
computations.  This check
leads us to suggest that for arbitray $M$ the formula for the
quantity
$F$ is given by
\begin{equation}
F(M)=Z_{WZNW}(-2c_V(G))~Z_{ghost}.\end{equation}
This can be justified by  nonperturbative computations at general
Dashen-Frishman
 conformal points \cite{Soloviev2}.

\bigskip

{\large\bf Appendix B}\vspace{.15in}
\renewcommand{\theequation}{B.\arabic{equation}}
\setcounter{equation}{0}

In this appendix, we
briefly give a first order  derivation  of the symmetry
(\ref{symmetry}) of the
action (\ref{action}) \cite{Soloviev6}.
First, we introduce auxiliary Lie-algebra-valued variables $Q, \bar
P$
and write the interacting
action in  the form
\begin{eqnarray}S(g_1,g_2,Q,\bar
P)&=&S_{WZNW}(g_1,k_1)~+~S_{WZNW}(g_2,k_2)\nonumber\\ & & \\
&-&{1\over2\pi}\int d^2z~\mbox{Tr}(k_1g^{-1}_1\partial g_1\bar
P+k_2Q\bar\partial g_2g_2^{_1}-{1\over2}Q\cdot S^{-1}\cdot\bar
P).\nonumber\end{eqnarray}
If we eliminate $Q, \bar P$ via their equations of motion, we regain
(\ref{action}). Instead, we express
$Q, \bar P$ in terms of new variables  $h_1,~h_2$ as
\begin{equation}
\bar\partial h_1h_1^{-1}=\bar P,~~~~~
h_2^{-1}\partial h_2=Q\end{equation}
and use the Polyakov-Wiegmann formula to obtain
\begin{eqnarray}S(g_1,g_2,Q,\bar
P)&=&S_{WZNW}(g_1h_1,k_1)~+~S_{WZNW}(h_2g_2,k_2)\nonumber\\ & &
\\&-&S_{WZNW}(h_1,k_1)~-~S_{WZNW}(h_2,k_2)
{}~+~{1\over4\pi}\int d^2z~\mbox{Tr}^2Q\cdot S^{-1}\cdot\bar
P.\nonumber
\end{eqnarray}
This is invariant under the transformations
\begin{equation}
g_1\to\bar\Omega_1(\bar z)g_1h_1\Omega_1(z)h_1^{-1},~~~~~~~~
g_2\to h_2^{-1}\bar\Omega_2(\bar z)h_2g_2\Omega_2(z),\end{equation}
as well as
\begin{equation}
h_1\to h_1\Lambda_1(z),~~~~~~~~
h_2\to  \bar\Lambda_2(\bar z)h_2.\end{equation}
The original model is regained on imposing
\begin{equation}
\bar\partial h_1h_1^{-1}=2k_2\mbox{Tr}~S~\bar\partial
g_2g_2^{-1},~~~~~
h_2^{-1}\partial h_2=2k_1\mbox{Tr}~S~g_1^{-1}\partial
g_1.\end{equation}

\bigskip

{\large\bf Appendix C}\vspace{.15in}
\renewcommand{\theequation}{C.\arabic{equation}}
\setcounter{equation}{0}

In this appendix, we shall discuss further the OPE given in eq.
(4.45).
The relevant operator $O^{L,\bar L}$ is the descendant at the first
affine
level of the adjoint-adjoint primary operator $\phi^{a\bar a}$. Let
us compute
first the OPE of $\phi^{a\bar a}$. For simplicity, we will consider
the special
case  $G=SU(2)$. Because of the holomorphic factorization, we can
present
$\phi^{a\bar a}(z,\bar z)$ as $\phi^a(z)\bar\phi^{\bar a}(\bar z)$.
Then we can
proceed with the left (holomorphic) part of $\phi^{a\bar a}$,
forgetting about
the right (antiholomorphic) part. Similar results can be derived for
the right
(antiholomorphic) part. Then
\begin{equation}
\phi^a(z)\phi^b(0)={[j=0]\over
z^{2\Delta_\phi}}~+~{C^{ab}_c\phi^c(0)\over
z^{\Delta_\phi}}~+~{[j=2]\over z^{2\Delta_\phi-\Delta_2}}~+~...
\end{equation}
Here $[j]$ are affine primaries with spin $j$ and conformal
dimensions
\begin{equation}
\Delta_j={j(j+1)\over k+2},~~~~~\Delta_\phi=\Delta_1.\end{equation}
$C^{ab}_c$ are structure constants.

For all positive $k$, $\Delta_\phi-\Delta_j<0$, when $j>1$.
Therefore, all
higher spin representations with $j\ge2$ do not contribute to the
singular part
of the OPE in eq. (C.1) \cite{Fateev}. However, we are interested in
$k<-2c_V$,
that is for $SU(2)$, $k<-4$. Hence, in this case all higher spin
primaries can
appear in the singular part of the OPE. Thus, it may seem as if all
these
singularities could appear in the OPE of the operator $O^L$ with
itsel
fmodifying the formula (4.45). For example, dimensional analysis
permits an
operator of spin 2 on the right handside of (4.45), so that
\begin{equation}
O^L(z)O^L(0)\sim{C_{[2]}[2]\over z^{2+2\Delta_\phi-\Delta_2}}~+~...,
\end{equation}
for some structure constant $C_{[2]}$.

Now we want to prove that the only singular terms which may appear in
the OPE
of two operators $O^L$ are given by eq. (4.45), so that in particular
$C_{[2]}=0$. Suppose that eq. (C.3) is correct. Acting with $J^a_0$
on both
sides of (C.3) gives
\begin{equation}
[J^a_0,O^L(z)]|O^L(0)\rangle~+~O^L(z)J^a_0|O^L(0)\rangle={C_{[2]}J^a_0
|[2]\rangle\over z^{2+2\Delta_\phi-\Delta_2}}~+~...\end{equation}
We have also acted on the $SL(2,C)$ vacuum, so that we deal with
states, not
operators.

Let us consider the case
\begin{equation}
L_{ab}\sim g_{ab}.\end{equation}
This matrix, for example, arises in eq. (4.46). Then it is easy to
show that
the left hand side of eq. (C.4) vanishes, since
\begin{equation}
J^a_0J^b_{-1}|\phi^b\rangle=0.\end{equation}
We have used the  relation
\begin{equation}
[J_n^a,\phi^b(z)]=\oint{dw\over2\pi
i}w^nJ^a(w)\phi^b(z)=z^nf^{ab}_c\phi^c(z).\end{equation}
This then implies the following condition
\begin{equation}
C_{[2]}~J^a_0|[2]\rangle=0.\end{equation}
Acting with $J^a_0$ and contracting indices, we obtain
\begin{equation}
C_{[2]}~g_{ab}J^a_0J^b_0|[2]\rangle=0,\end{equation}
but since $[j]$ has spin $j$, we also have
\begin{equation}
g_{ab}J^a_0J^b_0|[j]\rangle=j(j+1)|[j]\rangle,\end{equation}
which is non zero. Thus, in order for eq. (C.8) to be satisfied,
the coefficient $C_{[2]}$ must be zero,
\begin{equation}
C_{[2]}=0.
\end{equation}

Following the same chain of arguments, one can prove also that
\begin{equation}
C_{[j>2]}=0.
\end{equation}
This proves the OPE given by eq. (4.45) for the matrix $L$ in eq.
(C.5).

This result can be generalized to arbitrary matrices $L$ and
arbitrary groups
$G$ using the arguments given in section 4 and [28], but the above
constitutes
an explicit check for the special case $G=SU(2),~L_{ab}\sim g_{ab}$.

\end{document}